\documentclass[prd,nofootinbib,preprintnumbers,floats,aps]{revtex4}
\usepackage{graphicx,xspace,units}
\usepackage{float}
\usepackage{amsmath}
\usepackage{amssymb}
\usepackage[normalem]{ulem}
\usepackage{wrapfig}
\usepackage{epsfig}
\usepackage{bm}
\usepackage[usenames]{color}
\usepackage{ulem}

\newcommand{\eq}[1]{eq.~(\ref{#1})}
\newcommand{\Eq}[1]{Eq.~(\ref{#1})}
\begin{document}

\title{ Reflection above the barrier as tunneling in momentum space}
 
\author{R. L. Jaffe}
\email{jaffe@mit.edu}
\affiliation{Department of Physics,  Center for Theoretical Physics, and
Laboratory for Nuclear Science,
Massachusetts Institute of Technology,
Cambridge, MA 02139,
USA}

\begin{abstract}   
Quantum mechanics predicts an exponentially small probability that a particle with energy greater than the height of a potential barrier will nevertheless reflect from the barrier in violation of classical expectations.  This process can be regarded as tunneling in momentum space, leading to a simple derivation of the reflection probability.   
\end{abstract}
\maketitle


\section{Introduction} 
 
The most famous application of the semi-classical approximation in quantum mechanics is the calculation of the barrier penetration probability, $P(E)=\exp\left\{-\frac{2}{\hbar}\int_{a}^{b}dx\sqrt{2m(V(x)-E)}\right\}$, for a particle of energy $E$ to tunnel through a potential $V(x)$, when the height of the potential exceeds the particle's energy for $a<x<b$.
Elementary derivations of this result can be found in most textbooks\cite{griffiths}.  Quantum mechanics also predicts that a particle  with energy $E$, greater than a barrier $V(x)$ for all $x$, will reflect from the barrier.  This can also be estimated using semiclassical methods.  A derivation can be found in Landau and Lifschitz's well-known presentation of the semiclassical method\cite{Landaulifschitz}, but the derivation is more complicated and less intuitive than the derivation of the barrier penetration factor.  

In this paper we show that reflection above the barrier can be understood as barrier penetration \emph{in momentum space}, making the physical origin of the phenomenon more obvious and  reducing the derivation to that of ordinary barrier penetration.  This appears not to have been noticed in the past.  

An added benefit of this view of reflection above the barrier is that it leads to a simple derivation of the transition probability in the adiabatic approximation --- the Landau-Zener formula --- again based only on the physics of barrier penetration.

The paper is organized as follows: In Section II the problem of reflection above the barrier is shown to be barrier penetration in momentum space. In Section  III the result is shown to be identical to the  result derived in Ref.~\cite{Landaulifschitz}.  In Section IV a few  simple examples are presented.  In Section  V the transition probability in the adiabatic approximation is recast as reflection above the barrier in \emph{time} (as opposed to space) and the Landau-Zener result is derived.

\section{Tunneling in Momentum Space}

\begin{figure}
\begin{center}  
\includegraphics[width=8cm]{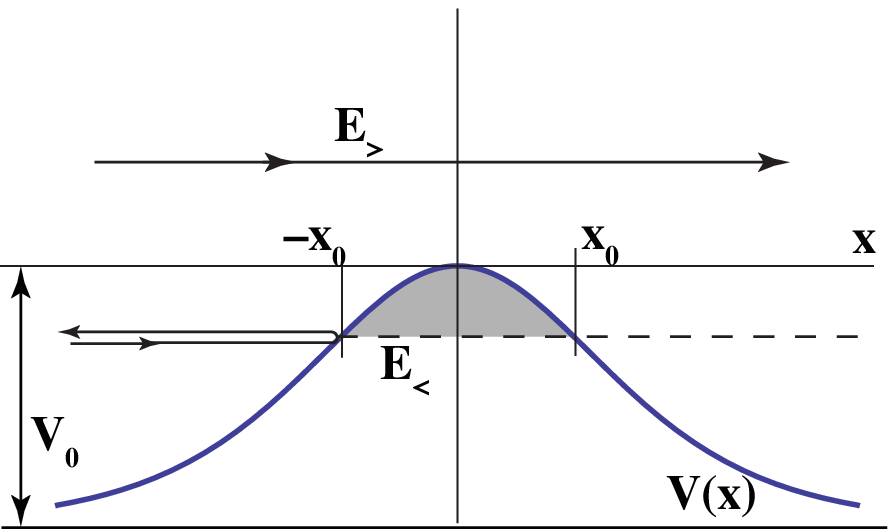}\hfill\includegraphics[width=8cm]{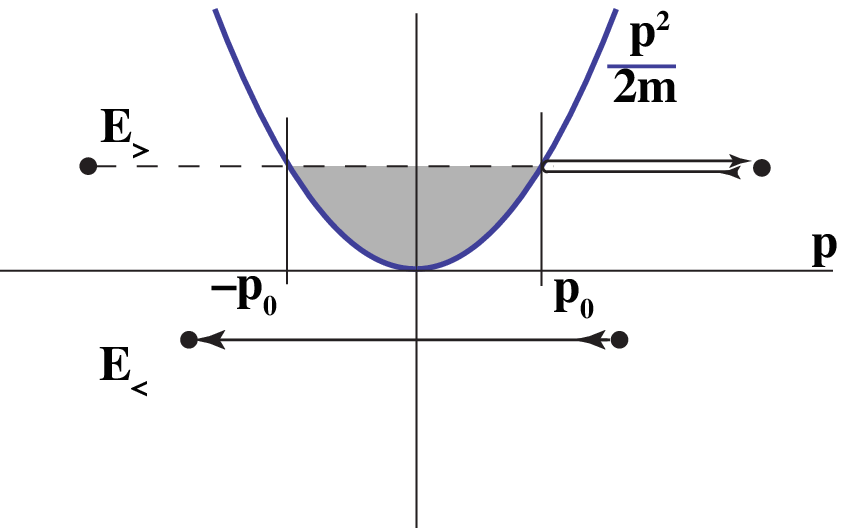}
\caption{\small Classical motion in the presence of a potential barrier in coordinate space (left panel) and momentum space (right panel).  On the left the classical particle either reflects from the barrier when $E=E_{<}$ is negative, or passes over it when $E=E_{>}$ is positive.  On the right the classical particle reflects from the (always parabolic) barrier when $E=E_{>}$ is positive and passes under it when $E=E_{<}$ is negative. }
\label{vofx}
\end{center}
\end{figure}

 To see  that reflection above the barrier can be viewed as barrier penetration in momentum space, consider the classical motion of a particle incident from the left on barrier $V(x)$, as viewed both in coordinate and   momentum space.  The familiar coordinate space depiction of the classical motion is shown in Fig.~\ref{vofx}(a).  For simplicity, take the barrier to be symmetric, $V(-x)=V(x)$, so $V_{\rm max}=V(0)$, and adjust the zero of energy so that $V_{\rm max}=0$.  The asymptotic value of the potential as $x\to \pm \infty$ is $-V_{0}$.   When $E$ is negative, for example $E_{<}$ in the figure, the particle moves up the~$x$-axis until it reaches the classical turning point at $x=-x_{0}(E_{<})$, where $V(-x_{0})=E_{<}$, then it turns around and retreats down the negative $x$-axis.  Quantum tunneling gives a small probability that the particle makes a transition to $x=+x_{0}(E_{<})$ and then propagates up the positive $x$-axis.  When $E$ is positive, for example $E_{>}$ in the figure, the particle propagates unimpeded up the $x$-axis from $-\infty$ to $+\infty$.

 Now   examine the problem from the point of view of momentum space, as shown in Fig.~\ref{vofx}(b).  For an energy $E$, the range of the momentum is bounded between $-p_{\rm max}$ and $+p_{\rm max}$, where $p_{\rm max}=\sqrt{2m(E+V_{0})}$, as marked by the circles on the trajectories in the figure. For positive energy, for example $E_{>}$, the particle  incident from the left in coordinate space  moves down the positive $p$-axis to the minimum classical momentum, $p_{0}(E_{>})=\sqrt{2mE_{>}}$, turns around, and retreats back up the positive $p$-axis.  This corresponds to the classically expected motion over the barrier in coordinate space. Reflection above the coordinate-space barrier occurs because quantum tunneling gives rise to a small probability that the particle makes a transition to $p=-p_{0}(E_{>})$ and then propagates on down the negative $p$-axis.   Having chosen $V_{\rm max}=0$, the classically forbidden region in momentum space always ranges from $-p_{0}(E_{>})$ to $p_{0}(E_{>})$ or {\it vice versa\/}.  When $E$ is negative, for example $E_{<}$ in the figure, the particle propagates unimpeded down the $p$ axis from $p_{\rm max}(E_{<})=\sqrt{2m(E_{<}+V_{0})}$ to $-p_{\rm max}(E_{<})$, corresponding to reflection from the barrier in coordinate space.
 
\begin{figure}
\begin{center}  
\includegraphics[width=10cm]{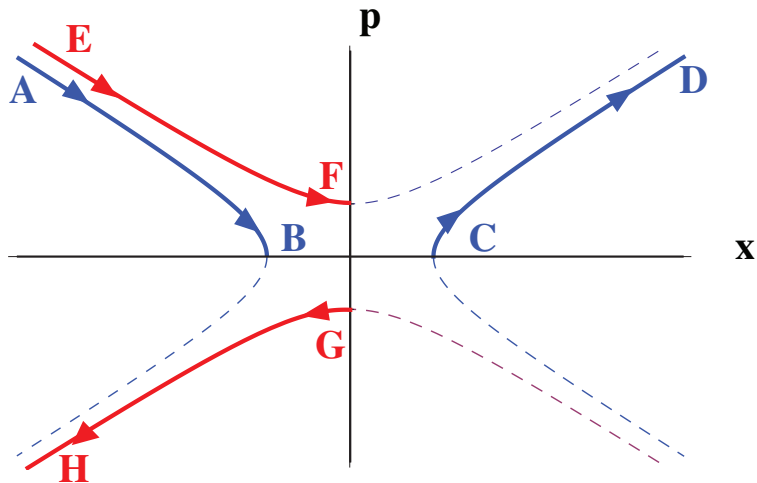}
\caption{\small Phase space trajectories corresponding to barrier penetration (in blue) ABCD and reflection above the barrier (in red) EFGH y for the harmonic oscillator.}
 \label{ho}
\end{center} 
\end{figure}

 The analogy is particularly clear for the case of a barrier that is an inverse harmonic oscillator, $V(x)=-\frac{1}{2}\alpha x^{2}$.  In that case the classical orbits are hyperbolae in phase space as shown in Fig.~\ref{ho}.  The barrier penetration trajectory ($E<0$) is shown in blue.  It progresses classically from $A$, at large negative $x$ and positive $p$, to the classical turning point $B$, Êthen jumps to $C$ and continues on to large positive $x$ and positive $p$.  The reflection-above-the-barrier trajectory ($E>0$) is shown in red.  It progresses classically from $E$, at large positive $p$ and negative $x$, to point $F$, jumps to $G$, and on to large negative $p$ and negative $x$.  The symmetry of the harmonic oscillator Hamiltonian under the exchange $p \leftrightarrow x$ (up to a change of scale) guarantees that both processes have analogous tunneling descriptions. 

To proceed quantitatively, consider 
 the Schr\"odinger equation for the momentum space wave function, $\phi(p)$,  
\begin{equation}
\label{momspace}
\left(\frac{p^{2}}{2m}+V(i\hbar\frac{d}{dp})\right)\phi(p)=E\phi(p)\,,
\end{equation}
where  $V$  has been written as a function of the operator $x\to i\hbar \frac{d}{dp}$, which is appropriate in momentum space.  Unless $V(x)$ is a sum of powers, $V(i\hbar \frac{d}{dp})$ is an integral operator\footnote{If $V(x)=\int  {dq}\tilde V(q)e^{-iqx/\hbar}$, then $V(i\hbar\frac{d}{dp})\phi(p)=\int {dq}  \tilde V(q)e^{q\frac{d}{dp}}\phi(p) =  \int dq \tilde V(q)\phi(p+q)$.
}. Things simplify significantly in the semiclassical approximation, where we expand the logarithm of the wave function in powers of $\hbar$,
\begin{equation}
\label{wkb}
\phi(p)=\exp \left(\frac{i}{\hbar}\sigma(p) +{\cal O}(\hbar^{0})\right)\,.
\end{equation} 
Substituting into eq.~(\ref{momspace}) and keeping the leading term in $\hbar$, we find\footnote{It is possible to compute the ${\cal O}(\hbar)$ term as well, but as in coordinate space, it does not affect the exponential barrier penetration factor.},
\begin{equation}
\label{wkbmom}
\frac{p^{2}}{2m}+V(x(p))=E\,, 
\end{equation}
where $x(p) \equiv -\sigma'(p)=- \frac{d\sigma}{dp}$.  
This equation can be solved for 
$\sigma(p)$ in terms of the inverse function $V^{-1}$,
\begin{equation}
\label{sigma}
\sigma(p)= \int^{p}dp'x(p') =- \int^{p}dp'\,V^{-1}(E-\frac{p'^{\,2}}{2m})
\end{equation}

In classically forbidden region of momentum space, $-\sqrt{2mE}<p<\sqrt{2mE}$, the inverse function, $V^{-1}(\xi)$ must be defined by analytic continuation from the domain $\xi<0$ where it is defined.  For example, in the case of the inverse harmonic oscillator, $V_{\rm HO}(x)=-\frac{1}{2}\alpha x^{2}$, $V^{-1}_{\rm HO}(\xi)=\pm\sqrt{-2\xi/\alpha}$, and
\begin{equation}
\label{invho}
V^{-1}_{\rm HO}(E-\frac{p^{2}}{2m}) = \pm\sqrt{\frac{2}{\alpha}\left(\frac{p^{2}}{2m}-E\right)}=\pm i \sqrt{\frac{2}{\alpha}\left(E-\frac{p^{2}}{2m}\right)}\,.
\end{equation}
Note that $V^{-1}$ is always imaginary in the classically forbidden zone.  The probability of reflection above the barrier, $|\mathfrak{R}(E)|^{2}$, is the exponentially small probability of barrier penetration in momentum space, which we obtain by  integrating $\int dp\, x(p) $ from $-p_{0}(E)$ to $p_{0}(E)$, and choosing the sign of $x(p)$ that corresponds to exponential suppression,
\begin{equation}
\label{reflectionfactor}
|\mathfrak{R}(E)|^{2}= \exp\left(-\frac{2}{\hbar}{\rm Im}\int_{-p_{0}(E)}^{p_{0}(E)}dp\ 
  V^{-1}\left(E-\frac{p^{2}}{2m}\right)   \right)\,,
\end{equation}
which is our result.

\section{Relation to  result of Landau and Lifschitz}

In their  treatment of the semi-classical approximation, Landau and Lifschitz\cite{attrib} derive a formula for the probability of reflection above the barrier that looks quite different from our \eq{reflectionfactor}.  They express the probability of reflection above the barrier as the exponential of a contour integral in the complex coordinate plane,
\begin{equation}
\label{LL}
|\mathfrak{R}_{\rm LL}(E)|^{2}=\exp\left(-\frac{4}{\hbar}\,{\rm Im}\,\int_{z_{1}}^{z_{0}}dz \,p(z)\right)\,, 
\end{equation}
where $p(z)=\sqrt{2m(E-V(z))}$.  $z_{0}$ is the \emph{complex} value of $z$ in the upper-half $z$-plane at which $p(z_{0})=0$, and $z_{1}$ can be any point on the real $z$-axis, since $p(z)$ is real everywhere on the real $z$-axis (remember $E>V(x)$ for all $x$).   For a symmetric potential with maximum at $x=0$, $z_{0}$ is on the imaginary $z$-axis, $z_{0}=iy_{0}$, so for convenience we take $z_{1}=0$, and rewrite \eq{LL} as
\begin{equation}
\label{LL2}
|\mathfrak{R}_{\rm LL}(E)|^{2}=\exp\left(-\frac{4}{\hbar}\,{\rm Im}\int_{0}^{y_{0}}i\,dy  \,p(iy)\right)\,.
\end{equation}
This result can be converted to ours using integration by parts,
\begin{equation}
\label{ibp}
\int_{0}^{y_{0}}dy\,p(iy)=\left.yp(iy)\right|^{y_{0}}_{0}+\int^{p_{0}}_{0}dp\,y(p)\,.
\end{equation}
The surface term vanishes because
$p(iy_{0})=0$ (This is how $z_{0}=iy_{0}$ was chosen.).  $p(0)$ is the classical momentum at $y=0$, so $p(y=0)=\sqrt{2mE}=p_{0}$, and the function $y(p)$ is the solution to \eq{wkbmom} with $x=iy$:   $p^{2}/2m+V(iy)=E$, whence   $y=-iV^{-1}(E-p^{2}/2m)$.  Substituting into \eq{LL2} converts it to our result, \eq{reflectionfactor}.  So, in essence, the relation between the two results is summarized by $\int pdx=-\int xdp$.  The derivation we have presented not only simplifies the mathematics,  but  also ``explains'' reflection above the barrier as barrier penetration in momentum space.
 
\section{Examples}

A few simple examples may make the application of \eq{reflectionfactor} more transparent.  We consider the inverse oscillator, a sech$^{2} x$ barrier, and a $(1 + x^{2})^{-1}$ barrier.
\subsubsection*{\bf{1.} $\mathbf{V_{\rm HO}(x)=-\frac{1}{2}\boldsymbol{\alpha} x^{2}}$}

The inverse of the harmonic oscillator potential is given by \eq{invho},
\begin{equation}
\label{harmonicinverse}
V_{\rm HO}^{-1}(E-\frac{p^{2}}{2m}) =   \frac{i}{\sqrt{\alpha}}\sqrt{2E-p^{2}/m}\,.
\end{equation}
Combining eqs.~(\ref{reflectionfactor}) and (\ref{harmonicinverse}) we obtain a reflection probability of
\begin{equation}
\label{harmonicbarrier}
|\mathfrak{R}_{\rm HO}(E)|^{2}= \exp\left(-\frac{2}{\hbar}\int_{-\sqrt{2mE}}^{\sqrt{2mE}}dp \sqrt{\frac{2}{\alpha}(E-\frac{p^{2}}{2m})}\right)=e^{-\frac{2\pi E}{\hbar\omega}}\,,
\end{equation}
where $\omega\equiv\sqrt{\alpha/m}$ is defined in  analogy with the frequency of a normal oscillator.  In this simple case, the exact result is known\cite{GY},
$$
|\mathfrak{R}_{\rm HO}(E)|^{2}_{\rm exact} = \frac{e^{-\frac{2\pi E}{\hbar\omega}}}{1+e^{-\frac{2\pi E}{\hbar\omega}}}\,,
$$
and agrees with the semiclassical estimate to leading order.

\subsubsection*{\bf{2.} $\mathbf {V_{2}(x)=V_{0}\,({ sech}^{2}(x/a)}\boldsymbol{-1})$}

In this case 
\begin{equation}
V_{2}^{-1}(E-p^{2}/2m)=\pm a\cosh^{-1}\left(\sqrt{ \frac{V_{0}}{E-p^{2}/2m+V_{0}}}\right).
\label{sechinverse}
\end{equation}
In the classically forbidden region, $E> p^{2}/2m$, and the argument of the $\cosh^{-1}$ is less than unity, where $\cosh^{-1}(y)=i\cos^{-1}y$.  Substituting into \eq{reflectionfactor}, we obtain the probability for reflection above the barrier,
\begin{align}
\label{sechbarrier}
|\mathfrak{R}_{2}(E)|^{2}&= \exp\left(-\frac{2a}{\hbar}\int_{-\sqrt{2mE}}^{\sqrt{2mE}}dp  
\cos^{-1}\left(\sqrt{\frac{V_{0}}{E- {p^{2}/2m}+V_{0}}}\right)\right)\nonumber\\
& = \exp\left(-\frac{2\pi a}{\hbar}\sqrt{2m}\left(\sqrt{E+V_{0}}-\sqrt{V_{0}}\right)\right)\,.
\end{align}
 
\subsubsection*{\bf{3.} $\mathbf{V_{3}(x) = -V_{0} x^{2}/(x^{2}+a^{2})}$}

In this case 
\begin{equation}
V_{3}^{-1}(E-p^{2}/2m) = ia\sqrt{\frac{E-p^{2}/2m}{V_{0}+E-p^{2}/2m}}\,,\,
\,\mbox{for} \,\,p^{2}<2mE\,,
\label{alginverse}
\end{equation}
and the reflection  coefficient can be expressed in terms of elliptic integrals,
\begin{equation}
|\mathfrak{R}(E)_{3}|^{2} =\exp\left(-\frac{4a}{\hbar}\sqrt{\frac{2mE}{1+\gamma}}\left((1+\gamma){\bf E}(\frac{1}{1+\gamma})-\gamma {\bf K}(\frac{1}{1+\gamma}\right)\right) \,,
\label{algbarrier}
\end{equation}
where $\gamma=V_{0}/E$ and ${\bf E}(x)$ and ${\bf K}(x)$ are complete elliptic integrals\cite{AandS}.  

As $E \to 0$, the probability of reflection grows and the sensitivity to the details of the shape of $V(x)$ disappears.  This can be seen explicitly by taking $E/V_{0}\ll 1$ in examples 2 and 3. In both cases the reflection probability reduces to the harmonic oscillator form with the parameter $\alpha$ replaced by the curvature at the top of the barrier, $\gamma\to 2V_{0}/a^{2}$ (the same for both the sech$^{2}$ and $1/(1+x^{2})$ barriers).

\section{The transition amplitude in the adiabatic approximation}

As a final remark, we use our approach to calculate the leading contribution to the transition amplitude in the adiabatic approximation.
According to the adiabatic theorem if a Hamiltonian, $H(t)$, changes slowly enough in time, a system initially prepared in an energy eigenstate will not make a transition to a different energy eigenstate, provided the two states remain well separated in energy.  If the Hamiltonian changes slowly and smoothly, violations of the adiabatic theorem  are exponentially small.  For the important case of two levels that come close to crossing,
the formula for the transition rate in the adiabatic approximation is  due to  Landau and Zener\cite{LZ}.   Qualitatively one can regard the Landau-Zener  transition as ``tunneling'' from one energy eigenstate to another.  In light of the ideas developed in this paper, it is therefore natural to seek a description of the transition process in the adiabatic approximation as \emph{reflection above the barrier} in the variable conjugate to energy, the \emph{time}.\footnote{This idea is not new, indeed in Ref.~\cite{Landaulifschitz} Landau and Lifschitz show that this problem is formally analogous to reflection above the barrier and obtain the transition probability by referring back to that problem.  The argument presented here is  quite similar but avoids the difficulties of analytic continuation to complex time.}

\begin{figure}
\begin{center}
\includegraphics[width=14cm]{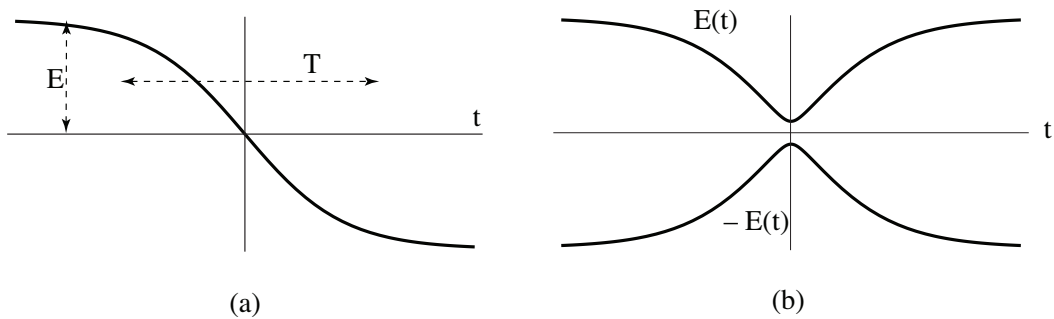}
\caption{\small (a) The function $f(t)$ that parameterizes the time dependence in the adiabatic approximation, and (b) The time dependent eigenvalues of $H(t)$. }
\label{foft}
\end{center}
\end{figure}

For simplicity I consider two states and a Hamiltonian\footnote{Without loss of generality we take $f(t)$ to be antisymmetric, $\epsilon$ to be real, and ${\rm Tr}\, H=0$.}
\begin{equation}
\label{h}
	H(t)=\left(\begin{matrix}
	f(t) & \epsilon\\ \epsilon & -f(t)
	\end{matrix}\right)\,,
\end{equation}
where $f(t)$ is a function that decreases monotonically with $t$ and is scaled by a parameter $T$, the ``transition time'', that determines the adabaticity.  We assume that 
$
\lim_{t\to\pm\infty}f(t)=\mp E\,,\,\mbox{with}\,\, E\gg \epsilon
$, so near $t=0$ the levels come close to crossing, but in fact repel.  
A typical form for $f(t)$ is shown in Fig.~\ref{foft}(a).

 At any time the eigenvalues of $H(t)$ are 
$$
E_{\pm}(t)=\pm\sqrt{f(t)^{2}+\epsilon^{2}}\equiv\pm E(t)\,,
$$ 
and the eigenvectors are
\begin{eqnarray}
\label{instsolns}
\phi_{+}(t)&=&\left(\begin{matrix} \cos\theta(t)/2 \\ \sin\theta(t)/2
\end{matrix}\right)\nonumber \\
\phi_{-}(t)&=&\left(\begin{matrix} -\sin\theta(t)/2 \\  \cos\theta(t)/2
\end{matrix}\right)\,,
\end{eqnarray}
where 
$$
\tan\theta(t) =  \frac{\epsilon}{f(t)}.
$$
A generic example of the energy level diagram is shown in Fig.~\ref{foft}(b).

In the adiabatic limit the full,   solutions to the time-dependent Schr\"odinger equation are simply proportional to $\phi_{\pm}(t)$,\footnote{We do not need to allow for an adiabatic phase here because we are not considering a \emph{closed} path in the space of states, and therefore the phase can be transformed away.} 
\begin{equation}
\label{psi+}
\Psi_{\pm}(t) =\phi_{\pm}(t)e^{\mp\frac{i}{\hbar}\int^{t}dt'E(t')}\,,
\end{equation}
  corresponding to $ E_{\pm}=\pm E(t)$.

Suppose that at large negative time the system is prepared in the eigenstate $\Psi_{+}$,
\begin{equation}
\Psi(t)\to \phi_{+}(-\infty)e^{-i\bar Et/\hbar}\,,\quad\mbox{as}\,\, t\to-\infty\, ,
\label{minus}
\end{equation}
where $\bar E=\sqrt{E^{2}+\epsilon^{2}}$. 
At large positive times the system will have a probability close to unity to remain in the state $\phi_{+}$, but there will be a small probability to find the system in the state $\phi_{-}$ in violation of the adiabatic approximation,
\begin{equation}
\Psi(t)\to \mathfrak{T}\, \phi_{+}( \infty)e^{-i\bar Et/\hbar} +\,\mathfrak{R} \,\phi_{-}(\infty)e^{+i\bar Et/\hbar}\,,\quad\mbox{as}\,\, t\to+\infty\,.
\label{plus}
\end{equation}
$|\mathfrak{R}|^{2}$ is the transition probability we seek to compute.  Note that eqs.~(\ref{minus}) and (\ref{plus}) define a problem analogous to scattering in one dimension.  Identifying $t$ with $x$, eqs.~(\ref{minus}) and (\ref{plus}) describe a  wave of amplitude $\mathfrak{T}$ incident from the right ($t\gg T$), giving rise to a transmitted wave of amplitude unity moving off to the left ($t \ll -T$) and a reflected wave of amplitude $\mathfrak{R}$ moving back to toward the right ($t\gg T$).  

To exploit this analogy we must convert the time dependent Schr\"odinger equation into a form that resembles the Schr\"odinger equation for scattering in one dimension.     First  write the  time dependent Schr\"odinger equation for this problem without approximation.  Let $\Psi(t)=\left(\begin{matrix}a(t)\\ b(t)\end{matrix}\right)$. Then $i\hbar\dot\Psi = H\Psi$ implies
\begin{eqnarray}
\label{firstorder}
i\hbar \dot a &=& fa+\epsilon b\nonumber \\
i\hbar \dot b &=& \epsilon a-fb.
\end{eqnarray}
By differentiating the first equation and substituting the second, we obtain a second order differential equation for $a(t)$,
\begin{equation}
\label{secondorder}
 -\hbar^{2} \ddot a - (f^{2}+\epsilon^{2})a-i\hbar\dot f a =0\,,
\end{equation}
 and an equation for $b(t)$ in terms of $a(t)$, 
 \begin{equation}
 \label{bfora} 
 b = \frac{1}{\epsilon}(i\hbar \dot a-fa)\,.
\end{equation}
The equation for $a(t)$ is a Schr\"odinger equation that describes scattering of a ``particle'' of mass $2m=1$ and energy $\epsilon^{2}$ from a ``potential''   
\begin{equation}
V(t) = - f^{2}(t) -i\hbar\dot f(t)\,.
\label{potential}
\end{equation} 
Eq.~(\ref{bfora}) merely fixes $b(t)$ given $a(t)$ and is irrelevant to our considerations.

When $T$ is large, $f(t)$ is slowly varying, and the semiclassical method applies.  The second term in $V(t)$ ($-i\hbar\dot f$) is higher order in $\hbar$ and can be ignored\footnote{This ``quantum'' contribution to the potential takes care of generating the prefactor $\cos\theta/2$ that modulates the semiclassical exponential in $a(t)$.}, and \eq{secondorder} describes motion of a zero energy particle \emph{above the potential barrier} $V(t)=-(f^{2}(t)+\epsilon^{2})$.  Without further ado, we can compute the transition rate by correspondence with   the arguments leading up to \eq{reflectionfactor},
\begin{eqnarray}
\label{subst}
V(x)&\to& - f^{2}(t)\,, \nonumber \\
E&\to& \epsilon^{2}\,,\\
2m&\to& 1\,.
\end{eqnarray}
So the transition probability in the semiclassical approximation is
\begin{equation}
\label{adabattrans}
|\mathfrak{R}|^{2}=  \exp\left(-\frac{2}{\hbar}\int_{-\epsilon}^{\epsilon}dp\ 
{\rm Im} \{f^{-1}(\sqrt{p^{2}-\epsilon^{2}}) \} \right)\,,
\end{equation}
which is our final result.  

\Eq{adabattrans} applies for an arbitrary slowly varying function $f(t)$.  The standard Landau-Zener formula applies to a \emph{linear level crossing}, where $f(t)=t/T$.  In this case $f^{-1}(\xi)=T\xi$ and  
\begin{equation}
|\mathfrak{R}|^{2}=  \exp\left(-\frac{4T}{\hbar}\int_{0}^{\epsilon}dp\ 
 \sqrt{\epsilon^{2}-p^{2}}\right) = \exp\left(-\frac{\pi T\epsilon^{2}}{\hbar}\right)\,,
 \label{LZresult}
\end{equation}
which is the Landau-Zener formula.  Of course \eq{adabattrans} applies to any slowly varying function of time.

\begin{acknowledgements}
I thank S. Jamal Rahi for a comments and suggestions.
 This work was supported in part by the U.S. Department of Energy under contract DE-FG03-92ER40701.
\end{acknowledgements}


\begin{thebibliography}{99}
\bibitem{attrib} Landau and Lifschitz attribute their result to Ref.~\cite{PSU}, but their derivation differs from that in Ref.~\cite{PSU}.
\end{thebibliography}
\end{document}